\documentclass[aps,12pt,nofootinbib,showkeys]{revtex4-1}
\usepackage[T1]{fontenc}
\usepackage[utf8]{inputenc}
\usepackage{amsmath}
\usepackage{graphicx}

\begin{document}

\title{Predictions on three-particle azimuthal correlations in proton-proton collisions}

\author{\c{S}ener \"{O}z\"{o}nder}
\email{ozonder@umn.edu}
\affiliation{Department of Physics, Istanbul Technical University, 34469 Istanbul, Turkey,}
\affiliation{Department of Physics, Ko\c{c} University, 34450 Istanbul, Turkey}
\date{\today}


\begin{abstract}
The ridge signal, which is long-ranged in rapidity, in the di-hadron
correlations in high-multiplicity p-p and p-A collisions opened up a whole new research area in high-energy QCD. Although the ridge had been observed in A-A collisions and
interpreted as a result of the radial flow of quark-gluon plasma,
it had not appeared until recently in the data of small collision systems
such as p-p and p-A, nor had it been predicted theoretically or seen
in the event generators. There are two competing approaches that attempt
to explain the systematics of the di-hadron ridge signal; hydrodynamics
and gluon saturation physics (color glass condensate/glasma). In this
work, we present predictions for the transverse momentum and rapidity
dependence of the three-particle correlation function within the gluon
saturation physics. Tri-hadron correlations can be measured, and the
data can possibly rule out one of the two alternative approaches. 
\end{abstract}

\maketitle

\section{Introduction}

In the relativistic proton (p) and nucleus (A) collisions, hundreds
of hadrons (pions, kaons etc.) and baryons (protons, neutrons etc.)
are produced, and the transverse momentum $p_{\perp}$,
rapidity $\eta$ and azimuthal angle $\phi$ of each particle are
measured. The correlations between the particles in rapidity and azimuthal
angle at a given transverse momentum interval often reveal interesting
information about the particle production mechanism in collisions as well as the
evolution of the interacting gas of particles until they freeze out
and reach the detectors.

The lowest-order correlation function is the two-particle correlation
function $C_{2}$, and it is a function of the azimuthal angle difference
$\Delta\phi=\phi_{1}-\phi_{2}$, rapidity difference $\Delta\eta=\eta_{1}-\eta_{2}$
and the two transverse momenta of gluon pairs. 
On the theory side, one calculates
the two-gluon correlation function, and convolves it with fragmentation functions
to obtain hadronic correlation function, which is measured at experiments.
One typically investigates
whether the particle pairs are correlated when their azimuthal angles
are at particular relative values such as $\phi_{1}\sim\phi_{2}$
and $\phi_{1}\sim\phi_{2}+\pi$, and whether these correlations are
preserved even while the rapidity difference $\Delta\eta$ between
the particles extends for several units. 

The high-multiplicity ($N_{\text{trk}}^{\text{offline}}\geq110$)
p-p collisions at the LHC at $\sqrt{s}=7\,\text{TeV}$ revealed
a very interesting phenomenon, which later was also seen in p-Pb collisions.
For the first time in p-p collisions, it has been observed that the strength
of the correlations between the produced hadrons at $\phi_{1}\sim\phi_{2}$
was preserved even when hadron pairs were separated for up to four units
of rapidity, $\Delta\eta\sim4$ \cite{Khachatryan:2010gv,Velicanu:2011zz,Li:2012hc,Wozniak:2017lgj}.
These correlations are called the ridge since they appear as a ridge when the
di-hadron correlation function plotted with respect to $\Delta \eta$ and $\Delta \phi$, 
and they have been known
to arise from radial expansion of the quark gluon plasma in A-A collisions
\cite{Dumitru:2008wn}. However, no such fluid flow or creation of
quark gluon plasma had been anticipated in p-p or p-Pb collisions.
Also, Monte Carlo event generators had not predicted emergence of
ridge correlations in these small systems where the target and projectile
overlap is small in comparison to that of A-A collisions. The ridge
correlations have been observed later in high multiplicity p-Pb collisions
as well \cite{Wozniak:2017lgj,CMS:2012qk,Abelev:2012ola,Aad:2012gla,Milano:2014qua,ABELEV:2013wsa,Abelev:2014mda,Abelev:2014mva,Aad:2014lta}.

In this work, we work in the framework of gluon saturation/glasma
framework, where no fluid flow or creation of quark gluon plasma is
necessary for emergence of the long-range azimuthal ridge correlations
\cite{Dusling:2009ni,Dusling:2009ar,Kovner:1995ts,Kovchegov:1997ke,Gelis:2009wh,McLerran:1993ni,McLerran:1993ka,McLerran:1994vd,Kovchegov:1996ty}.
The gluon saturation occurs when the density of gluons in a nucleon increases
and individual gluons start to overlap. At high energies, the gluon
density in the nucleon or nucleus becomes so high that one can think
of a nucleon or nucleus as a classical gluon field instead of individual
gluons\footnote{The density of quarks in the wavefunction of a nucleon or nucleus
is vanishingly small with respect to the density of gluons at high
energies. This can be seen through parton distribution functions of
quarks and gluons.}. It has been shown in Refs. \cite{Dumitru:2010iy,Dusling:2012iga,Dusling:2012cg,Dusling:2012wy,Dusling:2013oia,Venugopalan:2013cga}
that the glasma diagrams feautring gluon saturation gave rise to the
ridge correlations and they successfully explained the systematics of the
data, i.e., the change of the correlation strength with multiplicity,
$p_{\perp}$, $\Delta\phi$ and $\Delta\eta$.

The scale that gluon saturation becomes important and consequently
the classical gluon field approximation for the nucleon or nucleus
becomes justified is determined by the saturation scale. The saturation
scale increases with increasing beam energy $\sqrt{s}$, with which
also the multiplicity of produced hadrons increases. That the ridge
correlations only appear at high multiplicity events is an evidence
of the onset of gluon saturation. Also, the ridge correlations are
long-ranged in rapidity, and this can be explained in this framework where
the “running coupling Balitsky-Kovchegov” (rcBK) unintegrated gluon
distribution (UGD) functions are used to calculate the two particle
correlations. These UGDs feature gluon saturation and they have been calculated
by evolving the parton distributions functions in Bjorken-x 
variable\footnote{This is in contrast to the DGLAP evolution where the parton distribution
functions are evolved in $Q^{2}$.}.

Despite the success of the glasma model explaining the systematics
of the two-particle ridge signal, the debate whether the ridge arises
from gluon saturation or a small quark gluon plasma possibly formed
in the p-p and p-Pb collisions have not settled yet. In Refs. \cite{Ozonder:2014sra,Ozonder:2016xqn,Ozonder:2017moj},
we suggested that investigating the three-hadron correlations
could rule out one of the two mechanisms. 

In this work, we first show the formula for the three-gluon correlation
function $C_{3}$. Then we present our results on the three-particle
correlation function calculated with the rcBK wavefunctions. We also
make some predictions on the dependence of $C_{3}$ on $p_{\perp}$
and $y_{p}$ ($=\eta_{p}$) in p-p collisions at $\sqrt{s}=7\,\text{TeV}$.
These results can be compared with the data when $C_{3}$ is measured
in the future. 

\section{Three-particle azimuthal correlation function }

The triple-gluon inclusive distribution function is given by \cite{Ozonder:2014sra}

\begin{equation}
C_{3}=\frac{\alpha_{s}^{3}N_{c}^{3}S_{\perp}}{\pi^{12}(N_{c}^{2}-1)^{5}}\frac{1}{\boldsymbol{p}_{\perp}^{2}\boldsymbol{q}_{\perp}^{2}\boldsymbol{l}_{\perp}^{2}}\int\frac{d^{2}\boldsymbol{k}_{\perp}}{(2\pi)^{2}}({\cal T}_{1}+{\cal T}_{2}),
\end{equation}
where

\begin{align}
{\cal T}_{1} & =2\times\left(\Phi_{1,p}(\boldsymbol{k}_{\perp})\right)^{2}\Phi_{1,q}(\boldsymbol{k}_{\perp})\Phi_{2,p}(\boldsymbol{p}_{\perp}-\boldsymbol{k}_{\perp}){\cal T}_{A_{2}},\\
{\cal T}_{2} & =2\times\left(\Phi_{2,l}(\boldsymbol{k}_{\perp})\right)^{2}\Phi_{2,q}(\boldsymbol{k}_{\perp})\Phi_{1,p}(\boldsymbol{p}_{\perp}-\boldsymbol{k}_{\perp}){\cal T}_{A_{1}},
\end{align}
and

\begin{align}
{\cal T}_{A_{1}(A_{2})} & =\left[\Phi_{1(2),q}(\boldsymbol{q}_{\perp}-\boldsymbol{k}_{\perp})+\Phi_{1(2),q}(\boldsymbol{q}_{\perp}+\boldsymbol{k}_{\perp})\right]\\
 & \,\,\,\,\,\,\,\times\left[\Phi_{1(2),l}(\boldsymbol{l}_{\perp}-\boldsymbol{k}_{\perp})+\Phi_{1(2),l}(\boldsymbol{l}_{\perp}+\boldsymbol{k}_{\perp})\right].
\end{align}
Here $\alpha_{s}$ is the strong coupling constant, $N_{c}=3$ is
number of colors in QCD and $S_{\perp}$ is the overlap area of the
target and projectile. The first index of the UGD $\Phi$ is 1 or 2, and it refers to the projectile
and target. The second index of the UGD is the rapidity variable of
the produced gluon, and $\boldsymbol{p}_{\perp}$, $\boldsymbol{q}_{\perp}$
and $\boldsymbol{l}_{\perp}$ are the transverse momenta variables
of the gluons produced. 

To calculate the three-gluon correlation function, we use rcBK UGDs
\cite{Balitsky:1995ub,Kovchegov:1999yj,Dusling:2009ni,Gelis:2006tb,Fujii:2006ab}.
The details of evolution of these proton wavefunctions with rapidity
are given in Ref. \cite{Ozonder:2014sra}.

For three-gluons, there are two azimuthal angle differences ($\Delta\phi_{qp}=\phi_{q}-\phi_{p}$
and $\Delta\phi_{lp}=\phi_{l}-\phi_{p}$) and two rapidity differences
($\Delta\eta_{qp}=\eta_{q}-\eta_{p}$ and $\Delta\eta_{lp}=\eta_{l}-\eta_{p}$).
Also considering the magnitude of the transverse momenta of the three
gluons, the correlation function can be expressed as

\begin{equation}
C_{3}\equiv C_{3}(\Delta\phi_{qp},\Delta\phi_{lp},\Delta\eta_{qp},\Delta\eta_{lp},p_{\perp},q_{\perp},l_{\perp}).
\end{equation}

\begin{figure}
\begin{centering}
\includegraphics[scale=0.7]{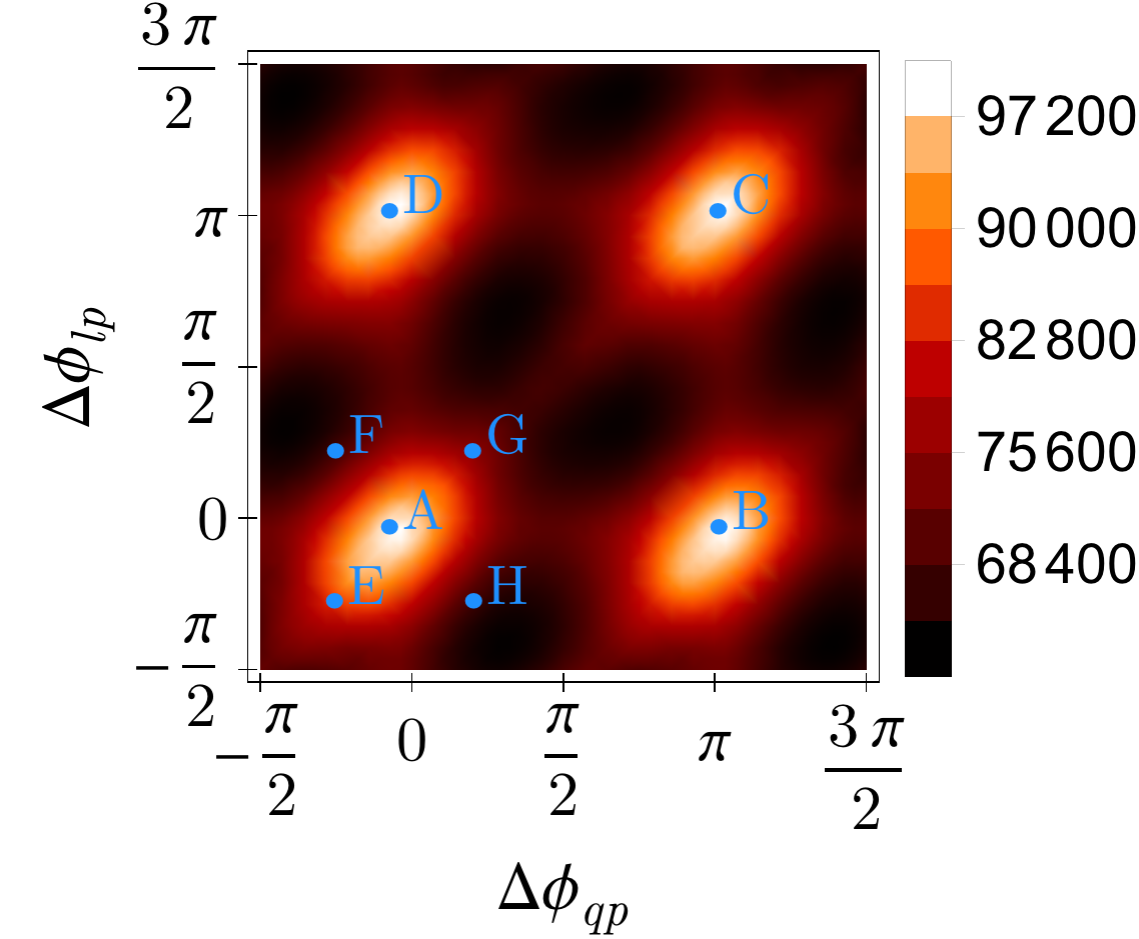} ~\includegraphics[scale=0.12]{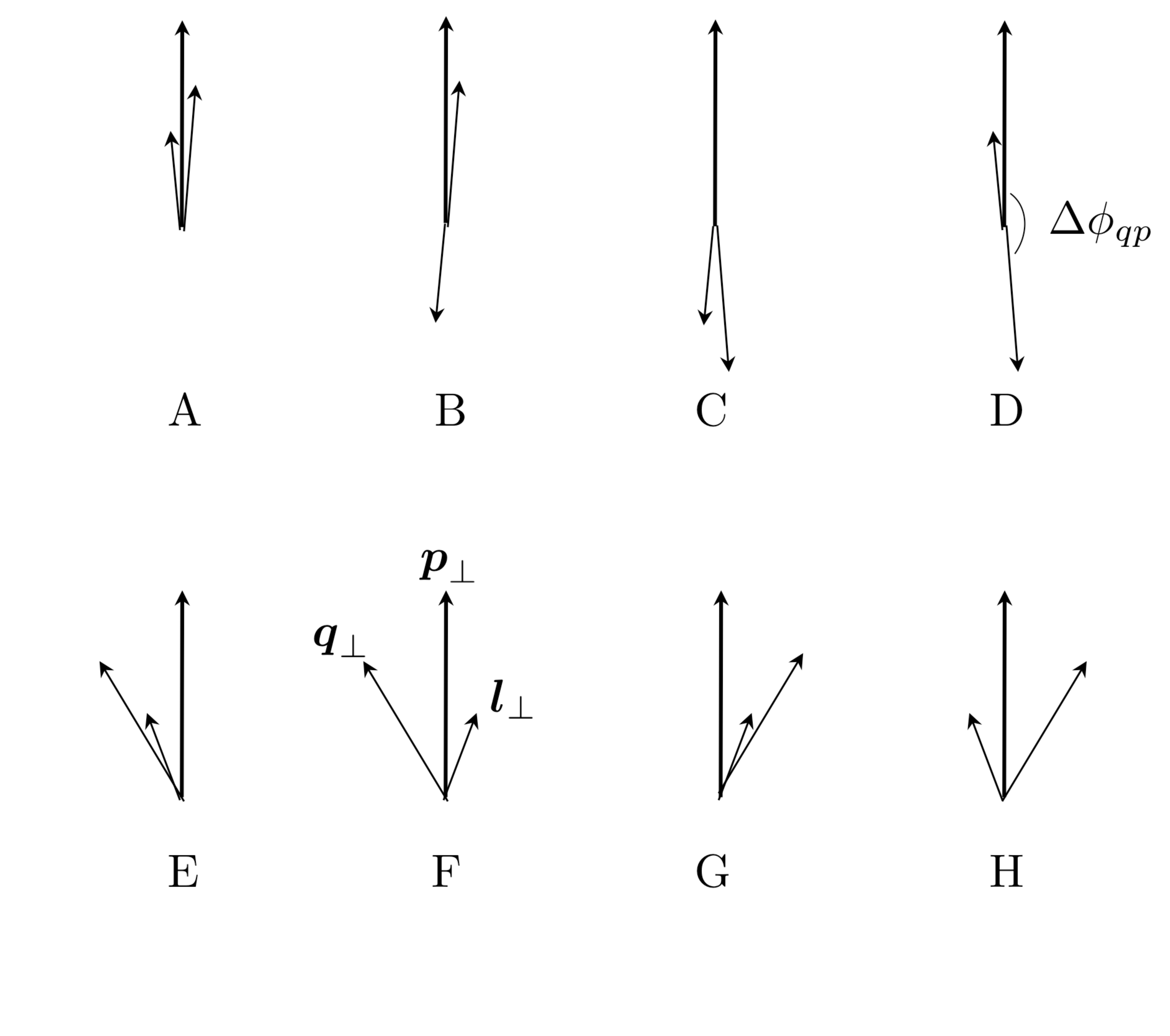}
\par\end{centering}
\caption{(left) The density plot of $C_{3}(\Delta\phi_{qp},\Delta\phi_{lp})$
in arbitrary units at transverse momenta $p_{\perp}=q_{\perp}=l_{\perp}=2\,\text{GeV}$
and rapidities $y_{p}=y_{q}=y_{l}=0$. (right) Here $p_{\perp}$ is
chosen as the trigger particle, so its azimuthal position is fixed.
The two azimuthal angle differences ($\Delta\phi_{qp}$ and $\Delta\phi_{lp}$)
are measured from the azimuthal position of gluon with momentum $p_{\perp}$.
This figure shows some possible azimuthal configurations of the three gluons
as marked with letters on the density plot. \cite{Ozonder:2014sra}}
\label{fig:densityplot}
\end{figure}

Figure \ref{fig:densityplot} shows a density plot of $C_{3}$ along
with several azimuthal configurations that different points on the
plot corresponds to.

\section{Systematics of three-gluon correlations}

In this section, we shall explore how $C_{3}$ changes with transverse
momentum and rapidity of the three gluons. In Ref. \cite{Ozonder:2014sra},
we showed that $C_{3}$ became maximum when $\{\Delta\phi_{qp},\Delta\phi_{lp}\}\approx\left\{ \left\{ 0,0\right\} ,\left\{ 0,\pi\right\} ,\left\{ \pi,0\right\} ,\left\{ \pi,\pi\right\} \right\} $.
This is in line with the finding that the peak in the two-particle ridge
correlations $C_{2}$ occured at $\Delta\phi_{qp}\approx0$ and $\Delta\phi_{qp}\approx\pi$
\cite{Dusling:2009ni,Dusling:2009ar,Kovner:1995ts,Kovchegov:1997ke,Gelis:2009wh,McLerran:1993ni,McLerran:1993ka,McLerran:1994vd,Kovchegov:1996ty}.
To study the momentum and rapidity dependence of $C_{3}$, here we
shall calculate it at one of its peak values with respect to the two
azimuthal angle differences; we arbitrarily choose this point to be
$\{\Delta\phi_{qp},\Delta\phi_{lp}\}=\left\{ 0,0\right\} $. This
point corresponds to the position marked as ``A'' in Fig. \ref{fig:densityplot}.

One of the hallmarks of the two-particle ridge correlations data is
that the magnitude of the peak at $\Delta\phi_{qp}\approx0$ is preserved
even if the two particles are separated by several units of rapidity
\cite{Dusling:2009ni,Dusling:2009ar,Kovner:1995ts,Kovchegov:1997ke,Gelis:2009wh,McLerran:1993ni,McLerran:1993ka,McLerran:1994vd,Kovchegov:1996ty}.
In the framework of saturation physics, this is attributed to the gluon
saturation and relatively slow evolution of the nucleon wavefunctions
with small-$x$ in the Bjorken limit of QCD. On the other hand, studies
show that application of hydrodynamics to p-p
and p-Pb collisions could also produce ridge correlations--without resorting
to gluon saturation--due to the assumed
hydrodynamization and consequent radial flow \cite{Bozek:2011if,Bozek:2012gr,Bozek:2013ska,Werner:2013ipa,Kozlov:2014fqa,Bzdak:2014dia}.
In order to better understand the actual mechanism behind how the
long-ranged rapidity correlations arise, examining the three-particle
correlations both in the saturation physics and hydrodynamics frameworks
is in order. Below, we show the dependence of $C_{3}$ on transverse
momentum and rapidity of one of the three gluons.

\begin{figure}
\begin{centering}
\includegraphics[scale=1.2]{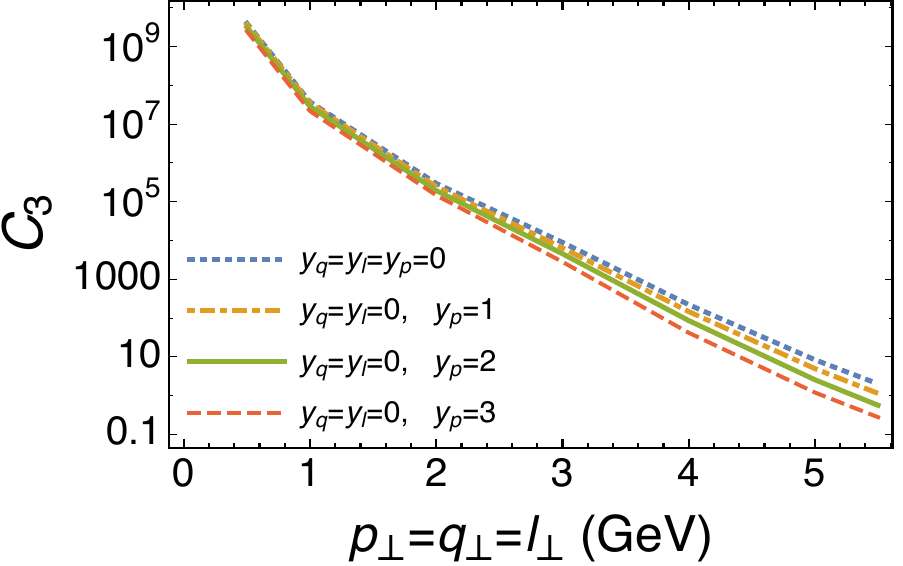}
\par\end{centering}
\caption{Log-linear plot of the three-gluon correlation function 
$C_{3}$ vs.~the transverse momenta of the gluons $p_{\perp}=q_{\perp}=l_{\perp}$
for p-p collisions at $\sqrt{s}=7\,\text{TeV}$. $C_{3}$ in the graph
is in units of $\alpha_{s}^{3}N_{c}^{3}S_{\perp}/\left[\pi^{12}(N_{c}^{2}-1)^{5}\right]$.
Different curves correspond to different rapidity values of the gluon
with the rapidity $y_{p}$.}
\label{fig:C3vspT}
\end{figure}

\begin{figure}
\begin{centering}
\includegraphics[scale=1.2]{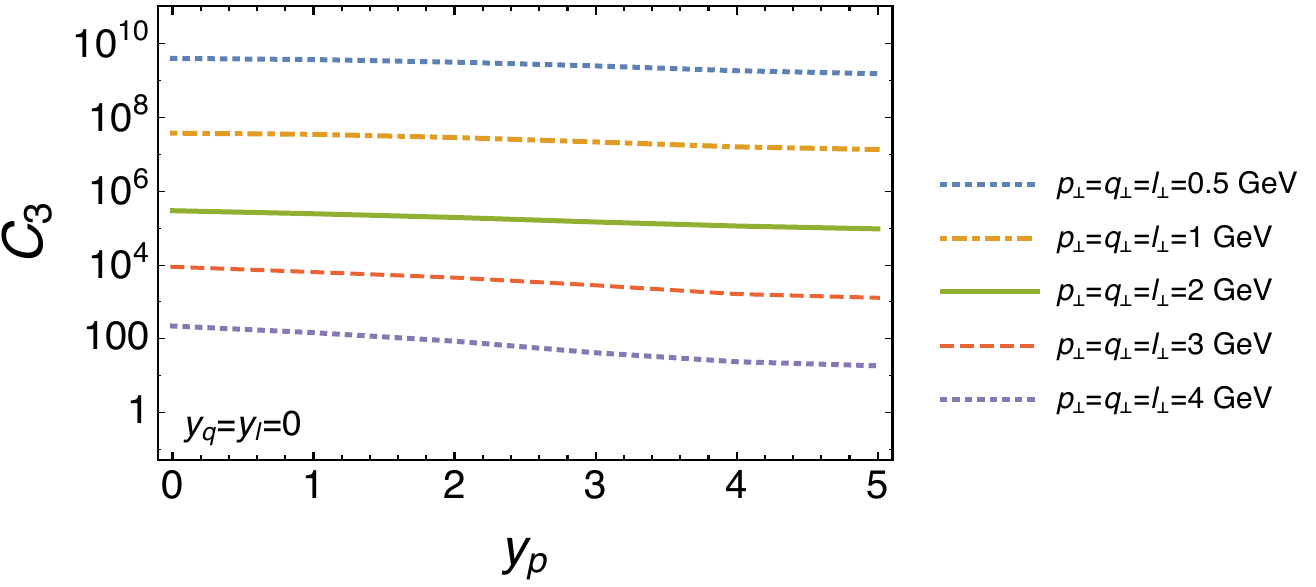}
\par\end{centering}
\caption{Log-linear plot of the three-gluon correlation function 
$C_{3}$ vs.~the rapidity $y_{p}$ for p-p collisions 
at $\sqrt{s}=7\,\text{TeV}$.
Here $y_{q}$ and $y_{l}$ are taken to be zero. $C_{3}$ in the graph
is in units of $\alpha_{s}^{3}N_{c}^{3}S_{\perp}/\left[\pi^{12}(N_{c}^{2}-1)^{5}\right]$.}
\label{fig:C3vsrapidity}
\end{figure}

The results of this work are shown in Fig. \ref{fig:C3vspT} and Fig.
\ref{fig:C3vsrapidity}. That the three-gluon correlations are large
at low-$p_{\perp}$ has also been observed in the two-particle correlation
calculations and measurements; this is understood as a manifestation
of the semi-hard gluon saturation scale, $Q_{s}\sim1\,\text{GeV}$
\cite{Dumitru:2010iy,Dusling:2012iga,Dusling:2012cg,Dusling:2012wy,Dusling:2013oia,Venugopalan:2013cga}.
As for the rapidity dependence of $C_{3}$, we observe that the correlation
strenght decreases approximately as power-law with increasing rapidity gap between the gluons. This is in contrast to the two-particle
correlations where the correlation strenght is approximately constant
with increasing rapidity difference.

\section{Summary and Outlook}

We presented results on the transverse momentum and rapidity
dependence of the three-gluon correlation function in the gluon saturation/glasma
framework. Our preliminary predictions are based on the gluonic correlation
function $C_{3}$, and in order to obtain the hadronic correlation
function, our results need to be convolved with the fragmentation function;
this is a subject of another work. However, we expect that the essential
features of our results based on the gluonic correlation function
would be preserved in the hadronic correlation function. The three-hadron
correlations in p-p collisions at $\sqrt{s}=7\,\text{GeV}$ have not
been measured yet. Our results can be compared with the data when
$C_{3}$ is measured in the future. We expect our work to be the next
step–after the study of di-hadron correlations– towards understanding
the true origins of the correlations in p-p and p-A collisions.

\section{Acknowledgement}

This work is funded by the Scientific and Technological Research Council of Turkey (TUBITAK) BIDEB 2232-117C008.

\bibliographystyle{apsrev4-1}    
\bibliography{three-particle-Turk-J-Phys}

\end{document}